# Optical isolation by temporal modulation: size, frequency, and power constraints.

Jacob B Khurgin

Johns Hopkins University, Baltimore MD 21208.

**Abstract**

Optical isolators are indispensable components of optical networks. Magneto-optic isolators have excellent operating characteristics, including low-to-no power consumption, but are not well suited for on-chip integration. The technique of temporal modulation of dielectric constant offers an alternative way to achieve isolation without magnetic field but is not without its own drawbacks. In this work I examine diverse methods of optical isolation via temporal modulation and show that independent on whether modulation is achieved by carrier injection, Pockels and acousto-optic effects, or any other conceivable method, there is essentially the same set of constraints on footprint, modulation frequency, and, most important, on power consumption required to achieve full isolation without excessive insertion loss. This power is estimated to be on the order of at least a hundred of milliwatts and whether this requirement is acceptable will depend on ongoing progress of both magneto-optic and time modulated integrated technologies.

- **Introduction**

Rather than sticking to the routine and commencing with a perfunctory narration of how indispensable are various optical devices in our lives, naturally due to relentless pursuit of boundless capacity for optical communications and other worthy goals, a narration which most readers would skip anyway, I will dispose with these formalities and just state that whether one likes it or not, but recent years have indeed seen accelerated (if not exponential) growth in the study of light propagation in time varying media[1]. Although most, if not all, of the phenomena associated with electromagnetic waves propagation in the medium with time-variable optical constants (chiefly permittivity) had been previously well understood as parametric interactions[2], the new millennium has seen a renewed interest in this field under new appellation, which in my view has grown organically from the field of metamaterials. In the development of metamaterials (and photonics crystals with which they often get confused) one forms materials with unconventional properties using engineering (or modulation) in space (2D or 3D) on sub-wavelength scale, so the desire to expand this engineering to the fourth, temporal dimension is logical[3]. And so, plenty of interesting works have appeared in the literature[4], including studies of time reflection and refraction[5], time crystals[6], realization of effective magnetic field[7] and synthetic dimensions[8], and last, but not least, investigation of nonreciprocity induced by time modulation. Nonreciprocity is important because in addition to being an inspiration for countless publications[9], it also underpins a real and essential functionality, optical isolation. Optical isolators (and circulators)[10, 11] are used to prevent reflections from reaching the laser and causing instabilities, to protect the detectors in lidars from direct laser radiation, and also to avoid parasitic feedbacks in communication links with optical amplifiers. Conventional optical isolators are based on Faraday effect in magneto-optic materials and perform their functions admiringly well, but they are relatively bulky and, when it comes to potential applications in photonic integrated circuits (PICS), they are not amenable to integration on silicon platform. That is why, in search of non-magnetic alternative to nonreciprocal devices, time modulated schemes gained prominence [1, 9, 12-14] with a tantalizing promise of compact integrated isolation.

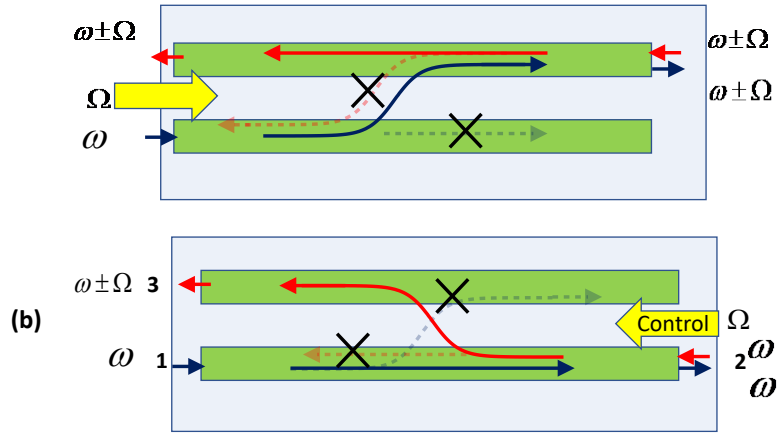

**Figure 1** General principle of optical isolation using time modulation. The schematically shown two "waveguides" coupled by the control wave modulating index can be two separate waveguides, two different modes in the same multimode waveguide, TE and TM waves in the same single mode waveguide, or simply the same mode with different frequencies. Mode separation in space is achieved by adiabatic coupling, polarization splitting, or, most often, narrow pass filter separating signals of different frequencies. The forward propagating optcal signal enters into Port 1 and Exits from Port 2. The backward propagating signal enters into Port 2 and routed into Port 3 (so the device can act as a circulator). In scheme (a) the forward signal is coupled into Port 2 by the forward propagating control wave of frequency while backward signal goes straight to Port 3 due to lack of phase (momentum) matching. Good isolation ratio is relatively easy to achieve but the insertion loss is high unless full coupling is achieved. In scheme (b) the forward signal propagates straight into Port 2 so the insertion loss is always low, but good isolation ratio is more difficult to achieve as it requires compete (or nearly complete) coupling.

In principle, the idea of isolation using time modulation is most easy to understand as a parametric process in which permittivity is modulated by a travelling wave (microwave[15], optical[16], acoustic [17-21], micromechanical motion[14, 22] and so on) and the process is phase (momentum) matched (or velocity synchronized) in forward direction and not phase matched in reverse. This is illustrated in Fig.1. Indeed, even before the current interest in time modulated propagation materialized isolation had been achieved using conventional traveling wave modulators [15, 23, 24] or simply a sequence of lumped electro-optic modulators [25, 26]. With all the progress achieved in recent years, the goal of developing non-magnetic isolators with performance comparable, if not superior to the magneto-optic isolators, still remains elusive. Commercial magnetic isolators[27] provide 40dB isolation ratio with negligible insertion loss (less than 1dB), and what is remarkable even if underappreciated fact, they consume no power. But these modulators are relatively long (few centimeters) and thus cannot be easily integrated into photonic circuits at this point, although the situation may change as there is a strong ongoing effort to develop integrated magneto-optic devices[28-30], which I briefly consider towards the end of this work. Thus, there exists a window of opportunity for integrated non-magnetic nonreciprocal devices and whether this opportunity will be successfully exploited depends on whether performance of the time-modulated devices can approach the system requirements in terms of high degree of isolation, low insertion loss, wide bandwidth, small footprint and, crucially, low power consumption. The latter is the main focus of the present work.

It is important to note here that in many time-varying schemes [16-21, 26] the forward propagating signal gets converted to a different frequency (sideband) and sometimes also into a different spatial mode, while backward propagating signal does not undergo conversion and is filtered away (Fig.1a) . Thus, a decent isolation can achieved no matter how low is conversion efficiency, and often very low operating power (less than a milliwatt [26]) can be claimed, but the effective insertion loss (ratio of output to input optical signal) then becomes unacceptably high (in excess of 20 dB). In other schemes [25, 31]the signal continues unconverted in forward direction while is depleted in backward direction (Fig.1b) hence it may in principle have 0dB insertion loss (or even parametric gain) but of course the power required for full conversion is much higher. Therefore, to provide a fair analysis of the required power for full isolation I shall adapt the power required for 100% conversion – then it does not matter which signal, converted, or not converted, is allowed to pass.

In this discourse (which is neither a review nor a tutorial) I adapt a rather uncommon inductive reasoning, starting with a specific time -modulated non-reciprocal scheme using Si waveguides –one of the few demonstrated experimentally[31]. Then, using it as a benchmark, I will show that no matter what active material is used in electro-optic modulation, the currents required for isolation are on the scale of few hundred milliamperes and the RF power on the order of 100 dBm (0.1 Watt). Furthermore, I establish that to achieve full isolation one has either to modulate at very high frequency, or, alternatively, settle for a large footprint. From that point, I will generalize the results to the arbitrary modulation method, be it electrical, acoustic or anything else as long as it relies on index change and establish that powers on the order of 100's of mw is required no matter which modulation method is adapted, unless, once again, one resorts to large effective device propagation length. The effective propagation length can be enhanced in the micro resonators, lowering the power requirement to tens of milliwatts but it comes at the expense of reduced bandwidth, increased complexity, not to count additional power required for maintaining the resonator tuned to laser wavelength. Towards the end, I will briefly review the recent advances in magnetic isolators and leave the readers (assuming their attention span allows them to reach this mark) to form their own opinion on practicality of using the time modulated nonreciprocal devices in current and future photonic circuits and networks.

- **Case Study: time modulation by carrier injection in silicon**

Consider the dispersion diagram of Fig.2a [12, 31]. Two propagating modes $E_i(x,y,z) = a_i(z)\hat{\mathbf{e}}_i(x,y)\exp(k_i z - \omega_i t) + c.c.$ where $a_i(z)$ are the amplitudes, $\hat{\mathbf{e}}_i(x,y)$ are the normalized field distributions, and $k_i$ are the propagation constants that follow dispersion laws $\omega_1(k_1)$ and $\omega_2(k_2)$ respectively. The group velocities $v_{g,i} = d\omega_i/dk_i$ can be equalized in order to assure the wide band operation. The low frequency perturbation $\delta\varepsilon(z)$ propagating in the forward direction can be written as $\delta\varepsilon(z) = \frac{1}{2}\Delta\varepsilon\exp[i(Kz - \Omega t)] + c.c.$. Change of permittivity can be engendered by propagating electro-magnetic or acoustic waves via electrooptic or acousto-optic effects. The electro-magnetic wave can simply freely propagate along some kind of transmission line (travelling wave modulation" ) or it can be emulated by judicious placing discrete electrodes at equal intervals and driving them with time-delayed replicas of the same harmonic signal. That is how it is usually achieved given large difference between propagation constants of two modes, $k_2(\omega_1+\Omega)-k_1(\omega_1)$ is much larger than wavevector of RF wave $K_{RF}=n_{RF}\Omega/c$ Then, when the forward phase matching condition $k_1(\omega_1)+K-k_2(\omega_1+\Omega)=0$ is satisfied a power transfer from mode 1 into mode 2 will result

$$\frac{da_2}{dz}=i\kappa a_1$$
$$\frac{da_1}{dz}=i\kappa a_2, \qquad (1)$$

where the coupling coefficient is $\kappa = \frac{\sqrt{\omega_1\omega_2}}{2n_{eff}c}\iint \mathbf{e}_1\cdot\Delta\varepsilon(x,y)\mathbf{e}_2^* dxdy$, and $n_{eff}$ is the effective index. With boundary condition $a_1(0)=a_0; a_2(0)=0;$ the solution for optical powers propagating in two modes is $a_1^2 = a_0^2\cos^2\kappa z$, $a_2^2 = a_0^2\sin^2\kappa z$, Therefore, if the length $L=\pi/2\kappa$ (half beat length) one achieves full transfer of power from mode 1 to mode 2.

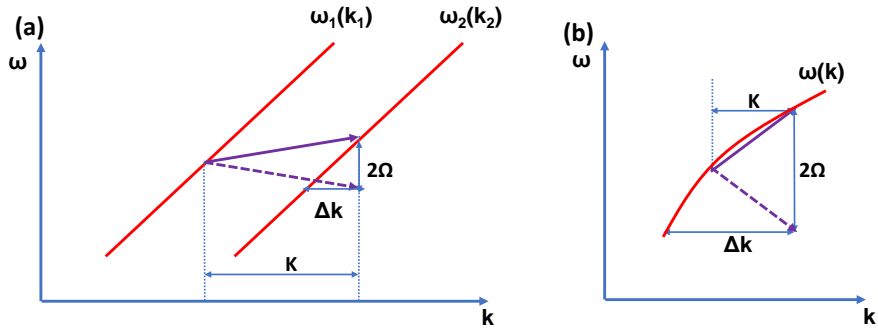

**Figure 2.** Momentum relations in the optical isolator (a) the coupling is allowed between two modes in one direction but impeded in the other direction by momentum mismatch $\Delta k$ (b) same for coupling within the same mode. $\Omega$ and $K$ are frequency and wavevector of the modulating wave.

When the light propagates in the opposite direction, the wavevector $K$ is pointed as shown as direction dashed arrow in Fig.2, hence the momentum mismatch between two waves is

$$\Delta k = k_1(\omega_1)+K-k_2(\omega_1-\Omega)=k_2(\omega_2+\Omega)-k_2(\omega_2-\Omega)=2\Omega/v_g \qquad (2)$$

Note that momentum mismatch is independent of the actual value of difference between the propagation constants of two modes, and, given that group velocity differs from the speed of light in vacuum by no more than a factor of a few, the only factor that affects $\Delta k$ is modulation frequency $\Omega$. Of course, the same mismatch occurs when the isolator uses conversion withing the same spatial mode as in Fig.2b

It is intuitively clear that to "resolve" the momentum mismatch $\Delta k$ the length should be at least $L \geq \pi/\Delta k = \Lambda/4n_g$, where $\Lambda = 2\pi c/\Omega$ is the RF wavelength in free space and $n_g = c/v_g$ is the group index. Indeed, in the reverse direction the coupled wave equations are

$$\frac{da_2}{dz} = i\kappa a_1 e^{i\Delta kz/2}$$
$$\frac{da_1}{dz} = i\kappa a_2 e^{-i\Delta kz/2},$$
(3)

with the solution of $a_2^2 = (\kappa^2/\gamma^2)a_0^2 \sin^2 \gamma z$, $a_1^2 = 1 - a_2^2$, where $\gamma = \sqrt{\kappa^2 + (\Delta k/2)^2}$. Therefore, to achieve full isolation, in which the forward propagating wave is completely depleted as power is transferred into mode 2, while the reverse propagating wave is not affected, one has to satisfy simultaneously two conditions, $L = \pi/2\kappa$ and $L = m\pi/\sqrt{\kappa^2 + (\Delta k/2)^2}$, where $m = 1,2...$ The shortest $L$ corresponds to $m = 1$ and therefore $\Delta k/2 = \sqrt{3}\kappa$, therefore $L_{min} = \pi\sqrt{3}/\Delta k = \sqrt{3}\Lambda/4n_g$. Therefore, in order to achieve isolation at reasonably short length one has to operate at relatively high frequency. Assuming a reasonable value of $n_g = 2$ one obtains the following expression for the product of modulation frequency $f = \Omega/2\pi$ and length $L$, $f \cdot L \geq 65 mm \cdot GHz$ so, for instance, modulating at 10GHz requires $L_{min} \sim 6.5mm$, which is significantly larger than the length of integrated magneto-optical isolators, If one desires to make the device more compact a higher modulation frequency is necessary, which, as we now show, inevitably increases power requirements. Note that this requirement for length essentially disappears if the modulating wave is optical with a very small, sub-micron wavelength, as in[16] , where a standard parametric convertor scheme was used. The required length is then determined only by the 100% conversion condition $L = \pi/2\kappa$. But such scheme requires presence of a second laser with high power.

Next, consider the power/current requirements. The 100% isolation occurs when , $\kappa = \pi/2L$, and if the change of refractive index is $\Delta n$, we obtain $\Delta n F_{12} L/\lambda = 1/4$, where the coupling overlap factor is $F_{12} = \iint \mathbf{e}_1 \cdot \Delta\varepsilon(x,y) \mathbf{e}_2^* dxdy / \Delta\varepsilon_{max} < 1$. Let us first consider the most explored case where the mechanism of index change is carrier injection, as, for example in Si[32], where the index change can be estimated as $dn/dN \sim .8 \times 10^{-21} cm^3$, and $N$ is the density of carriers (electrons and holes). If the active ross-section of the area into which the carriers are injected is $S_{act}$, the total charge injected is

$$Q_{inj} = eS_{act}L\Delta N = eS_{act}(dn/dN)^{-1}\lambda/4F_{12} = eS_{eff}(dn/dN)^{-1}\lambda/4 \qquad (4)$$

, where $S_{eff} = S_{act}/F_{12}$ is on the scale of a few times the mode area. The current required to achieve isolation is $I = \Omega Q_{inj} = e\Omega S_{eff}(dn/dN)^{-1}\lambda/4$. For $S_{eff} \sim 0.1\mu m^2$, $\lambda = 1.55\mu m$ and 10GHz modulation frequency one obtains $I = .5A$. That is a very significant current, at 10GHz frequency. As far power goes, if the injection is achieved in p-n junction, the voltage swing is on the scale of a few hundred mV, and the RF power required to achieve isolation is a few hundred milliwatts. Indeed, in the experimental implementation with 25dBm (300 mw) only less than 3dB isolation had been achieved ,so one can surmise that closer to 30dBm (1W) of power would be needed to achieve full isolation. To reduce footprint of the devices, say to 2mm, would require modulation at 30GHz and even higher power on the scale of 3W which would be a very difficult task. Using alternative method of carrier injection in MOSFET-like configurations[33] in place of p-n junction, would require even higher gate voltages to achieve the same carrier density modulation (typically a few volts) and that in turn would lead to increase in power requirements.

- **Going beyond silicon**

And what about using other materials instead of Si? The last few years have seen a flurry of reports of new materials used for electro-optic modulation, among them transparent conductive oxides, such as ITO (indium tin oxide)[34], and two-dimensional materials, such as graphene[35] and transition metals di-chalcogenides (TMDC)[36]. All of these modulators, just like Si rely on carrier injection as means for index modulation and can be referred as "charge driven" [37, 38] modulation schemes. In ITO, just like in Si, modulation is achieved by changing plasma dispersion, in graphene it is state filling (moving the Fermi level) that does the trick, and in TMDC's the modulation is the result of exciton bleaching, but, despite all these apparent differences, the net modulation efficiency is roughly comparable for all of these methods[38]. It follows, from the simple fact that the degree by which each injected carrier can change the refractive index depends on the value of the dipole matrix element of the optical transition (or effective mass, as the two are related) and on detuning from the resonance. For an allowed transition in a given region of spectrum the value of a dipole is restricted by the oscillator sum rule and amounts to 1-2 Angstrom (bond length) and the detuning has to be fairly large to avoid excessive absorption.

To apply the knowledge gained in the comparative analysis of electrooptic modulator[38] to the task of optical isolation, one should note that the isolator of [31] looks exactly like a MZI modulator with the only difference that the waveguides in two arms are coupled to each other. In fact, in MZI modulator applying voltage (injecting current) to one of the waveguides accomplishes transfer from symmetric to anti-symmetric mode, and, moreover the whole process of intermodal frequency conversion amounts to a carrier-suppressed single sideband modulation that can be implemented using MZI modulators [39].

If one considers a simple electro-optic modulation in a single mode waveguide the amount of charge required to achieve 180 degrees shift can be determined by the same expression (4) but with a different form of overlap factor $F_{11} = \int\int \mathbf{e}_1 \mathbf{e}_1^* \cdot \Delta\varepsilon(x,y) dx dy / \Delta\varepsilon_{max}$, i.e. a simple overlap of the active index change region with the mode. Therefore, one can use the main result developed in [38] for the electro-optic modulator (Eq.21 there), describing the charge required to achieve 180 degrees,

$$Q_\pi = en_{eff} S_{eff} / \sigma_\Phi \tag{5}$$

where $\sigma_\Phi$ is the phase shift cross section (essentially a phase shift per unit density of carriers) that has been estimated for various current-injection modulation schemes and found to be about $10^{-16} cm^2$ for ITO and MoS$_2$ and three times as much for the case of graphene (due to high Fermi velocity). We can therefore estimate the current required to achieve the isolation using (5) (there are two factors differentiating the isolation from simple modulation – only $\pi/2$ phase change is needed but on the other hand only one half (positive frequency) of index modulation is used – so these factors cancel each other) and obtain $I = en_{eff}\Omega S_{eff}/\sigma_\Phi$. Using the same parameters as for Si modulator above we obtain current required for complete isolation of about 300mA for graphene and almost an Ampere for other modulators, i.e., power on the order of a Watt.

To summarize, isolating using with current driven modulator using dielectric waveguides requires RF power on the order of 100's of mw. and lengths of a few mm, no matter what the active material is. The length requirement makes use of plasmonic waveguides with smaller effective cross section unrealistic since waveguide losses over few mm of propagation length will be enormous. One has to revert to ring geometries to increase the effective length, but that solution is also far from being perfect as discussed towards the end of this work

**Other means of electro-optic modulation and advantages of traveling wave modulation**

It should be noted, that with all the attention given to charge driven modulators promising the miniaturization, no decent isolation in them has been demonstrated as of yet. At the same time, using III-V [15, 23] and LiNbO$_3$ [24, 25] Pockels modulators for isolation purposes has been demonstrated on many occasions. Broadly speaking, Pockels modulators as well as one based on Quantum Confined Stark Effect, can be characterized as field-driven [37] in which change of refractive index is proportional to the electric field $\Delta n = n^3 rE/2$, where $r$ is electro-optic coefficient. Nevertheless, one can still introduce the switching charge, by noting the relation between the charge density on the electrodes and electric field, $E = F_{EO}Q/\varepsilon_0 \varepsilon wL$, where $w$ is the width of the electrode and $F_{EO} < 1$ describes overlap between the RF field distribution and optical mode. Then one can arrive at the following expression for switching charge on the electrode, similar to (4)

$$Q_{elec} = 2\varepsilon_0 \varepsilon w\lambda / 4rn^3 F_{EO} = 2\varepsilon_0 \varepsilon S_{eff} \lambda / 4rn^3 t_m = eS_{eff}(dn/dN_{eff})^{-1}\lambda/4, \qquad (6)$$

where $dn/dN_{eff} = ern^3 t_m / 2\varepsilon_0 \varepsilon$ denotes the change of refractive index in the waveguide with the change of effective charge density in the waveguide (i.e., the total charge on the electrodes divided by volume of the waveguide. Note the presence of $t_m$-effective height of the mode in the waveguide in this expression, which signifies the main difference between charge and field driven modulation. In charge driven scheme each injected electron only changes the refractive index locally, while in a field driven scheme each carrier added to electrodes changes the refractive index throughout entire waveguide. Thus, it is no wonder that for LiNbO$_3$ (r=30pm/V, n=2.2, static dielectric constant averaged over the directions $\varepsilon = 55$, $t_m = 400nm$) one obtains $dn/dN_{eff} \sim 0.5 \times 10^{-20} cm^{-3}$ which is a few times higher than what one gets for any carrier injection driven modulator. This fact, as well as lower loss explains why despite the progress in novel carrier injection modulators, field driven modulators like LiNbO$_3$ still retain their dominance in practical applications[40]. Therefore, using LiNbO$_3$ in lumped configuration one may reduce the required current to maybe 100mA and the power to a bit less than 100mW. Similar results can probbaly be obtained with InP and GaAs modulators based either on Franz Keldysh s or Quantum confined Stark effect (QCSE)[41]

Now, if one uses field driven modulators, one can use traveling wave rather than lumped modulators which significantly reduces the current and power required for isolation. As shown in Supporting Information, when one uses traveling wave, the charge propagates from one end of modulator to another, rather than gets injected in and out every cycle. As a result, the current and power both get reduced by a factor $\Omega \tau_{tr} = 2\pi n_{RF} L / \Lambda$, where $\Lambda / n_{RF} = 2\pi / K_{RF}$ is the microwave wavelength in the medium. Since $K_{RF}$ is small one has to use two modes with relatively close propagation constants, for instance the fundamental TE and TM modes with an appropriate EO coefficient (for example $r_{51}$ providing their coupling). Essentially, one only needs to charge the capacitor with length $\Lambda/2\pi n_{RF}$ to start the wave propagation along the microwave waveguide – thus the current and power can be reduced in traveling wave isolator, but at the expense of the larger length. Since for 10GHz modulation the wavelength $2\pi/\Lambda \sim 0.5 cm^{-1}$ significant reduction of power would require length in excess of 1cm. Of course, one also has to consider microwave losses in the electrodes that will reduce this advantage, especially if one attempts to increase the frequency. Then the only feasible way appears to be to resort to ring resonators, as will be discussed further down this text.

- **What power would it take to isolate via time modulation by any other conceivable means?**

As a next, and hopefully final step towards obtaining the general picture of what it takes to obtain nearly perfect (30dB or better) isolation with any other time modulated scheme, whether driven electrically, optically, acoustically, or by any other means. First thing to note is that the range of refractive indexes in the optical range is remarkably narrow, from 1.4 to 2.4, and to maybe 3.5 if one extends the range to short wave infrared[42, 43]. To have significantly different indices two materials have to have completely different densities, and lattice structure. For example, Si and GaAs are two very different materials from many points of view (e.g. one is direct and the other indirect semiconductor with different mobilities) but their refractive indices are almost identical. This is quite different from the microwave region [44], where the (static) dielectric constants range from 4 in $SiO_2$ to almost 100 in $TiO_2$ to almost 2000 in ferroelectrics like $SrTiO_3$ (corresponding indexes ranging from 2 to 10 to 45). This fact implies that static dielectric constant can be tuned with a relative ease in comparison to that in the optical range. In addition, simple injection/depletion of carriers (varactor) allows tuning of the capacitance (and therefore effective index of microwave waveguide) by orders of magnitude. It is not a wonder then that all the time modulation feats from isolators [13] to time crystals [45] have been successfully demonstrated in microwave range, while in optical range the results have been less impressive[46]

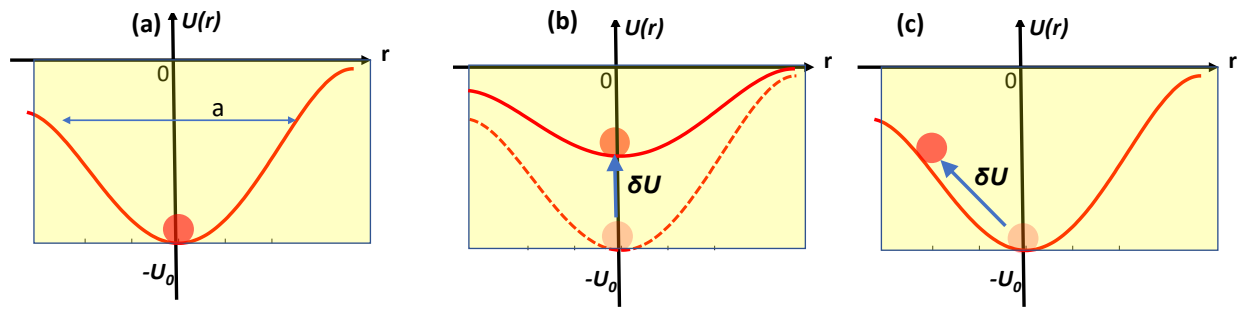

**Figure 3** (a) Simple model of the origin of polarizability and dielectric constant in optical range – valence electron oscillating in the anharmonic potential. Dielectric constant is determined by the curvature of the potential. Polarizability and dielectric constant can be changed by (b) reducing the depth of the potential, or (c) by promoting the electron to higher energy where the curvature of the potential is reduced due to anharmonicity.

So let us now assume a very simple picture of optical(dielectric) response of an anharmonic oscillator (Fig 3a) in which the potential $U(x) \sim -U_0 + \frac{1}{2}Kx^2 + ...$ where $U_0$ is some kind of "binding" or "cohesive" energy and the $K \sim 2U_0/a^2$, where $a$ is the spatial extent of the bound state. The order of magnitude of $U_0$ is a few eV and its width $a$ is commensurate with a bond length, i.e. a couple of angstroms indicating that $K \sim 10^{20} - 10^{21} eV/m^2$, and the resonant frequency (in energy units) is $\hbar\omega_0 \sim \sqrt{\hbar^2 K/m_0} \sim 3-8 eV$. This resonant frequency in solids corresponds to the average bandgap, often referred to as Penn gap[47] (which is always significantly larger than fundamental bandgap). In this simple model the dielectric constant in the transparency region well below $\omega_0$ but above all the phonon resonances can be modeled as $\varepsilon \sim 1 + Ne^2/\varepsilon_0 m_0 \omega_0^2 \sim 1 + Ne^2 a^2/\varepsilon_0 U_0$ where $N$ is the density of valence electrons which is proportional to $a^{-3}$. Now, if one is to change the dielectric constant by the amount $\delta\varepsilon/\varepsilon \sim 1$ one has to change the energy of each individual valence electron by $\delta U/U_0 \sim 1$. One way this can be accomplished is by changing the depth of potential (hence it's curvature K), which is what happens in ion-related part of electro-optic effect, acousto-optic effect, or thermal nonlinearity and schematically shown in Fig.3b.. The other way, shown in Fig.3c, is by promoting each electron to the energy state close to 0 (once again changing K due to anharmonicity), which occurs in electronic part of electro-optic effect (Pockels or Kerr), in nonlinear optical effects of various orders, as well as by removing all N carriers from the valence band (the latter is equivalent to injecting N carriers into the valence band). Note that each of these operations implies reduction of the total binding energy to a very small value, i.e. causing break down. This is of course exactly what is observed in experiments: as relative change of permittivity approaches 100% (and usually long before it) a complete breakdown of the material ensues.

Now, all the phenomena leading to change of refractive index depend on the energy density inside the material. If one considers Pockels effect or second order optical nonlinearity the change is proportional to the electric field (RF or optical), i.e. square root of the energy density $u$ $\Delta n/n \sim (u/NU_0)^{1/2}$. With $N \sim 10^{23} cm^{-3}$ one obtains $NU_0 \sim 10^4 - 10^5 J \cdot cm^{-3}$. Using Pockels effect in LiNbO$_3$ [48] as an example we obtain $\Delta n/n \sim n^2 rE/2 \sim (u/u_0)^{1/2}$, where $r_{33} = 30 pm/V$ is Pockels coefficient $u = \varepsilon_0 \varepsilon E^2/2$, $\varepsilon = 30$; is static dielectric constant, and $u_0 = 2\varepsilon_0 \varepsilon/n^4 r^2 \sim 2.5 \times 10^4 J/cm^3$. To change refractive index by applying optical rather than radio frequency wave one can use second order nonlinearity of GaAs[49] with $\chi^{(2)} \sim 180 pm/V$ and obtain $u_0 = 2\varepsilon_0 n^4/|\chi^{(2)}|^2 \sim 8 \times 10^4 J/cm^3$. For organic chromophores Pockels coefficients are larger on the order of $r \sim 100 pm/V$ [50] leading to a much lower $u_0 \sim 10^3 J/cm^3$ - that can be explained by lower density of organic molecules compared to the density of valence electron in crystals and also due to low effective biding energy $U_0$ - the latter resulting in thermal instability of the organic matter which largely prevented organic modulators from widespread practical applications despite high Pockels and nonlinear coefficients.

The same can be said about acousto-optic effect [51], where index change is proportional to the amount of strain $S$ and can be written as $\Delta n/n \sim n^2 PS/2$, where $P$ is a relevant photoelastic tensor element, and energy density is $u = \rho v_s^2 S^2$, where $\rho$ is the mass density and $v_s$ is the speed of sound. Introducing acousto-optic figure of merit $M_2 = n^6 P^2/\rho v_s^3$ we obtain once again $\Delta n/n \sim (u/u_0)^{1/2}$, where this time $u_0 = 2n^2/M_2 v_s$. Using one of the better acousto-optic materials, TeO$_2$ ( $M_2 = 35 \times 10^{-15} s^3/kg$, $n = 2.2$, $v_s = 4,250 m/s$ as an example, we obtain $u_0 \sim 7 \times 10^4 J/cm^3$. Therefore, one can generalize the same dependence $\Delta n/n \sim (u/u_0)^{1/2}$, where $u$ is the energy density injected and stored in the material (not necessarily absorbed!) and $u_0 = NU_0 > 2 \times 10^4 J/cm^3$ holds for any effect that is linear with the applied force. Then, using condition $\Delta n F_{12} L/\lambda = 1/4$ we obtain

$$P_{is} = (\lambda/4n)^2 L^{-1} S_{eff} u_0 \Omega \tag{7}$$

and if we assume the length equal to minimum length required for isolation, $L_{min} = 2\pi\sqrt{3}c/4\Omega n_g$ we can re-write this equation as

$$P_{is} = \pi\sqrt{3}(\lambda/4nL)^2 S_{eff} u_0 c/n_g \tag{8}$$

Just to gain order of magnitude idea of the power required for full isolation we take, $L = 3mm$, $S_{eff} = 2\mu m^2$, and $u_0 = 2 \times 10^4 J/cm^3$, and obtain $P_{is} \sim 100mW$. Note that this result was obtained assuming no additional loss of efficiency due to, for example, microwave loss in the electrodes, coupling loss and as well as reduced mode overlap factors.

For Kerr effect (RF or optical) and thermal nonlinearity the change is proportional to the energy density itself, $\Delta n/n \sim (u/NU_0)$. Hence

$$P_{is} = (\lambda/4n) S_{eff} u_0 \Omega \tag{9}$$

i.e. a factor $4nL/\lambda = (4/2\pi)\omega\tau_t$, larger than (7), where $\tau_t$ is the propagation time, Since typically this propagation factor is at least 2 orders of magnitude speaks in favor of using phenomena with square root dependence of index modulation, $\Delta n/n \sim (u/u_0)^{1/2}$ i.e. linear electro-optic, acousto-optic, or second order nonlinear effects.

One should also acknowledge a new class of modulators, based on doped transparent conductive oxide with permittivity (real part) approaching zero at a given frequency – the so-called "epsilon near zero (ENZ) materials, such as for instance ITO (Indium Tin Oxide)[52, 53] or AZO (Aluminum Zinc Oxide)[54]. In these materials one can obtain large relative change of permittivity or index. The permittivity in ENZ materials has two parts, $\varepsilon = \varepsilon_b - \chi_f$, where $\varepsilon_b$ is the permittivity due to bound carriers (interband) and $\chi_f$ is the susceptibility of the free carriers (intraband) that nearly balance each other. Modulation of index is achieved by injecting the electrons into conduction band or by heating them optically within the band taking advantage of band nonparabolicity[55, 56]. For this reason, in the expression $\Delta n/n \sim (u/u_0)^{1/2}$ one should use $n' \sim \sqrt{|\chi_f|} \sim \sqrt{\varepsilon_b}$ which is on the scale of 2-3 as in any dielectric rather than very small $n = \sqrt{\varepsilon}$ in order to obtain the right magnitude of the index change. Thus even though the relative change of index can be enormous, the absolute change is no different from any other modulation based on carrier injection. At any rate, the propagation distance in ENZ materials is very short due to high loss, so t full optical isolation cannot be achieved in them.

- **Using ring resonators**

Since as it seems that even under the most favorable conditions the power required to achieve a full isolation in non-resonant geometries is on the scale of at least 100mW and the length is at least a few millimeters, one way to improve performance is to resort to resonant scheme, typically involving mirroring (or whispering gallery) resonators as in [13, 17-19, 26]. One can make a simple estimate of improvement by noting that the effective interaction length in the resonator $L_{eff}$ in resonator is increased by a factor $1/\alpha_{rt}$ relative to the ring perimeter $L$, where $\alpha_{rt}$ is the round trip loss. Then in place of (7) one obtains

$$P_{is} = (\pi\lambda/2Fn)^2 L^{-1}\alpha_{rt}^2 S_{eff} u_0 \Omega \tag{10}$$

Now using $\alpha_{rt} = 0.03$ and $L = 1mm$ (less than 0.3mm² footprint and 10GHz modulation frequency we obtain isolating power of about 25mW. But one has to consider the fact that additional power will be expended in order to keep the ring resonance tuned to the laser frequency, which will be at the very least 20mW [57, 58]. When one considers the fact that the modulated wave must by physically separated from the unmodulated one, on both ends of the isolator, additional filters are required that also need to be precisely tuned, increasing both footprint and power consumption. Therefore, even with ring resonators one expects power consumption required for full isolation to be on the scale of 100mW.

- **State of the art of the integrated magnetic isolators**

Having established the power, frequency, and length requirements for time modulation driven nonreciprocity one is expected to make a comparison with the existing magnetic isolators, where the development also had not been frozen in place over the last decade. That I will do, reluctantly, as I my goal is not to question of rationality of the relentless pursuit of nonmagnetic resonators, but simply to When it comes to magnetic isolators, conventional wisdom of difficulty of using them in photonic integrated circuits no longer remains unchallenged as numerous integrated magneto-optic devices have been developed, using the in-plane magnetic fields and TM waves in asymmetric waveguides incorporating magneto-optic layers such as Ce:YIG. [28, 29, 59, 60]. With TE polarization the situation is somewhat more difficult, but here one should note that all time modulated schemes in the literature are also not polarization independent .The footprint of magnetic isolators according to[60] can be as small as 1x0.3 mm, isolation ratio as high as 30dB and insertion loss as low as 5/10dB for TM/TE polarization. Since MZI had been used in these modulators the bandwidth is very wide. When it comes to power requirements one has to note that usually coyly omitted fact that permanent magnets consume exactly zero power which may make time-modulated schemes a bit less attractive. However, there is a line of thought according to which the presence of permanent magnet is undesirable since it is difficult to integrate. But integrated magneto-optic modulators can also be made without permanent magnet, using current flowing through microstrip to generate magnetic fields sufficient for 30dB isolation [59] A typical value of current required for full isolation is only 100-200mA which is far less than in time modulated schemes, and the power is less 10mW, i.e., roughly at least an of magnitude less than in time modulated isolators. So, to summarize this part, integrated magnetic isolators have smaller footprint, lower insertion loss, consume orders of magnitude less power and require no microwave circuitry, and, unlike integrated time modulation schemes have been successfully demonstrated (success meaning isolation in excess of 20dB) on more than one occasion. Whether these obvious advantages are outweighed by the constraints associated with purported difficulties of integration on Si platform is not for me to say – the decision should be made by application engineers.

- **Parting thoughts**

As I have already mentioned, this work is neither a comprehensive review nor a tutorial, but simply an attempt to estimate the power, size, and frequency requirements for achieving optical isolation without magnetic field. The combination of size and frequency requirements for isolation leads to the necessity of modulating at frequencies exceeding 10GHz. The necessity of achieving change of refractive index $\Delta n$ in the material is shown to require presence of energy $u_0 \Delta n / n$ per unit volume in the medium, where $u_0 \sim 10^4 - 10^5 J/cm^3$, no matter what modulation technique is used: electro-optic, all-optical, acousto-optic, opto-mechanical, and so on. As a consequence, full isolation in the reasonably short device with a few mm length requires powers in excess of 100 mw, not counting inefficiencies associated with propagation and coupling losses. Reduction of power can be achieved by several means. One can increase the length of the device and use traveling wave geometry (but large footprint would be a disadvantage). One may also attempt to reduce the effective cross-section of the device, using plasmonics, but high loss associated with metal would restrict the length and make full isolation unattainable. Finally, the most promising isolation method involves multiple micro rings, allowing reduction of power to tens of milliwatts, but the additional power required for tuning the micro-resonators and also filtering the frequency shifted output would bring the power requirements back to 100mW. Whether expending few hundred milliwatts to achieve optical isolation of single laser device is an acceptable price for getting rid of magnetic materials with much lower power requirements (ideally zero) is the question that does not have an easy answer and will depend on future developments of both magneto-optic and time modulated isolating technologies.

**Funding**

This project was funded by the Air Force Office of Scientific Research (FA9550-16-10362) and DARPA Defense Sciences Office (HR00111720032).

**Acknowledgements**

I would like to acknowledge generous support by DARPA DSO in the framework of now expired NLM program that for four years allowed me to indulge in unrestrained and unsupervised research activities leading to this contribution. And as always, the unwavering and resolute support given to me by my dear colleagues Prof. P. Noir and Dr. S. Artois has been one factor that had let me to carry this work to its conclusion.

Supporting Information Available: **Advantages of traveling wave isolators** This material is available free of charge via the Internet at http://pubs.acs.org

Supporting Information for

# Optical isolation by temporal modulation: size, frequency, and power constraints.

## Advantages of traveling wave isolators

Consider a field-driven lumped electro-optic modulator with the total charge on electrodes required for isolation $Q_{elec}$ (Equation 6 in the main text). For simplicity we consider a simple plane parallel capacitor geometry with thickness $t$, length $L$ and width $w$. The electric field is $E = Q_{elec} / \varepsilon_0 \varepsilon w L$ hence the energy density is $u = \frac{1}{2}\varepsilon_0 \varepsilon E^2$, where $\varepsilon$ is the dielectric constant at RF frequency, and the total isolating energy is

$$U = \frac{1}{2}\frac{Q_{elec}^2 t}{\varepsilon L w} = \frac{1}{2}\frac{Q_{elec}^2}{C} \tag{1}$$

where capacitance $C = \varepsilon_0 \varepsilon w l / t$. The power consumption of the lumped device is

$$P_{lump} = U\Omega = \frac{1}{2}\frac{Q_{elec}^2 t \omega}{\varepsilon_0 \varepsilon L w} = \frac{1}{2}\frac{Q_{elec}^2}{C}\Omega, \tag{2}$$

and since the current is $I = Q_{elec}\Omega$, the impedance then can be found as $|Z| = t / \Omega \varepsilon_0 \varepsilon L w = 1/\Omega C$

Now, if we have a traveling wave device the power consumption can be found via the RF power density flowing through the waveguide

$$P_{tw} = \frac{1}{2}\frac{wtE^2 n}{\eta_0} = \frac{1}{2}\frac{tQ_{elec}^2 n}{\eta_0 \varepsilon_0^2 \varepsilon^2 L^2 w} = \frac{1}{2}\frac{Q_{elec}^2 n}{\eta_0 \varepsilon_0 n^2 L C} = \frac{1}{2}\frac{v}{L}\frac{Q_{elec}^2}{C}, \tag{3}$$

where $n_{RF} = \sqrt{\varepsilon}$, and $v = 1/n_{RF}\sqrt{\varepsilon_0 \mu_0} = 1/n_{RF}\varepsilon_0 \eta_0$ is the microwave propagation velocity So, the ratio of power consumption for the traveling wave and lumped isolators is

$$\frac{P_{tw}}{P_{lump}} = \frac{v}{\Omega L} = \frac{\Lambda}{2\pi n_{RF} L} = \frac{1}{\Omega \tau_{tr}}, \tag{4}$$

where $\tau_{tr} = L/v$ is the transit time. Thus the total power gets reduced compared to the lumped modulator. Obviously, the total current is also reduced by the same ratio, and is equal to $I = Q_{elec}/\tau_{tr}$, while impedance is increased to

$$Z_{tw} = (t/w)\eta_0 / n, \tag{5}$$

and is frequency independent.